\documentclass[10pt]{article}
\usepackage{fullpage}
\usepackage{amsmath}
\usepackage{amssymb}
\usepackage[dvips]{epsfig}
\usepackage{color}

\def\##1{{\bf #1}}
\def\=#1{\underline{\underline #1}}

\def\eps{\epsilon}

\def\.{\mbox{ \tiny{$^\bullet$} }}

\def\le{\left(}
\def\ri{\right)}

\def\lec{\left\{}
\def\ric{\right\}}

\def\c#1{\cite{#1}}
\def\l#1{\label{#1}}
\def\r#1{(\ref{#1})}

\begin{document}

\begin{center}
\Large{\bf {\LARGE On limitations of the  Bruggeman formalism for
inverse homogenization }}

\normalsize \vspace{6mm}

Siti S. Jamaian and Tom G. Mackay\footnote{Also affiliated to:
NanoMM~---~Nanoengineered Metamaterials Group, Department of
Engineering Science and Mechanics, Pennsylvania State University,
University Park, PA 16802--6812, USA. Email: T.Mackay@ed.ac.uk}

\vspace{4mm}

\noindent{ \emph{School of Mathematics  and
   Maxwell Institute for Mathematical Sciences\\  University of Edinburgh,
 Edinburgh EH9 3JZ,
United Kingdom} }

\vspace{12mm}

%%% Abstract
{\bf Abstract}\end{center}

The Bruggeman formalism provides an estimate $\eps^{Br}_{hcm}$ of
the relative permittivity of a homogenized composite material (HCM),
arising from two component materials with relative permittivities
$\eps_a$ and $\eps_b$. It can be inverted to provide an estimate of
$\eps_a$, from a knowledge of $\eps^{Br}_{hcm}$ and $\eps_b$.
Numerical studies show that the inverse Bruggeman estimate $\eps_a$
can be physically implausible when (i) $\mbox{Re} \lec
\eps^{Br}_{hcm} \ric / \mbox{Re} \lec \eps_b \ric > 0 $ and the
degree of HCM dissipation is moderate or greater; or (ii) $\mbox{Re}
\lec \eps^{Br}_{hcm} \ric /$ $ \mbox{Re} \lec \eps_b \ric$ $ < 0 $
regardless of the degree of HCM dissipation. Furthermore, even when
the inverse Bruggeman estimate is not obviously implausible, huge
discrepancies can exist between this estimate and the corresponding
estimate provided by the inverse Maxwell Garnett formalism.
\vspace{5mm}

\noindent {\bf Keywords:} Bruggeman  formalism; Maxwell Garnett
formalism; inverse homogenization; metamaterials

\vspace{5mm}

\section{Introduction}

A composite material may be regarded as being effectively
homogeneous, provided that wavelengths  are much larger than the
particle sizes of the component materials that make up the composite
material. The constitutive parameters of such a homogenized
composite material (HCM) can be estimated from a knowledge of the
constitutive parameters of its component materials, along with a
knowledge of the distributional statistics and shapes of its
component particles \c{L96,M_electromagnetics}. The Bruggeman
homogenization formalism has been widely applied for this purpose
for the past 70 years \c{L96,Ward}; and new areas of application for
the Bruggeman formalism continue to emerge, for examples, in recent
developments pertaining to complex HCMs \c{WM_AEU,MW_JOPA} and
negatively--refracting metamaterials
\c{M_electromagnetics,ML_NPV_MOTL}. However, a certain limitation of
the Bruggeman homogenization formalism came to light in 2004
\c{ML_OC1}. In the context of isotropic dielectric HCMs, derived
from two component materials with relative permittivities $\eps_a$
and $\eps_b$, it transpires that the Bruggeman estimate of the HCM
relative permittivity may be physically implausible if $\eps_a /
\eps_b < 0$ in the case of nondissipative HCMs or if $\mbox{Re} \lec
\eps_a \ric / \mbox{Re} \lec \eps_b \ric < 0 $ in the case of weakly
dissipative HCMs.
 An
example of such a problematic homogenization scenario~---~of
considerable interest to the metamaterial community~---~arises in
the homogenization of silver particles with insulating particles at
visible and near infrared wavelengths \c{M_Ag}. This
limitation~---~which is relevant to active \c{ML_OC2} as well as
nondissipative and dissipative HCMs~---~also extends to the Maxwell
Garnett homogenization formalism which shares a common provenance
with the Bruggeman formalism \c{Aspnes}, as well as the
Hashin--Shtrikman, Wiener and Bergman--Milton bounds on the HCM's
relative permittivity \c{ML_OC1,Duncan}. Similar anomalous results,
for HCMs arising from  two component materials with $\mbox{Re} \lec
\eps_a \ric / \mbox{Re} \lec \eps_b \ric < 0 $, have been described
as `electrostatic resonances' \c{Bross1,Bross2,Bross3}, but  this
term is avoided in \c{ML_OC1,M_Ag,ML_OC2,Duncan} (and herein) since
the estimates of the HCM's relative permittivity  described in
\c{ML_OC1,M_Ag,ML_OC2,Duncan} are not physically plausible.

Restricting our attention to the simplest possible case of an
isotropic dielectric HCM arising from two isotropic dielectric
component materials, in this communication we investigate the
applicability of the Bruggeman  formalism  to the inverse
homogenization scenario wherein the relative permittivity of one of
the component materials is estimated from a knowledge the relative
permittivities of the other component material and the HCM. Formal
expressions have been established for the inverse Bruggeman
formalism (and the inverse Maxwell Garnett formalism) in the general
setting of bianisotropic HCMs \c{WSW_MOTL}, but in certain cases
these formal expressions may be ill--defined \c{Cherkaev} and the
ranges of applicability of these inverse formalisms have not been
established. Our study is partly motivated by very recent
implementations of the inverse Bruggeman formalism in estimating
nanoscale constitutive and morphological parameters of certain
sculptured thin films \c{ML_JNP}, which is a key step in modelling
the electromagnetic response of infiltrated sculptured thin films
\c{ML_PJ,ML_PNFA}.

\section{Analysis and numerical studies}

We consider the homogenization of two isotropic dielectric component
materials with relative permittivities $\eps_a$ and $\eps_b$. The
component materials $a$ and $b$ are assumed to be distributed
randomly as spherical particles with volume fractions $f_a$ and
$f_b= 1- f_a$, respectively.
 The Bruggeman estimate of the
relative permittivity  of the corresponding HCM, namely
$\eps^{Br}_{hcm}$, is provided via \c{Ward}
\begin{equation}
f_a \frac{\eps_a - \eps^{Br}_{hcm}}{\eps_a + 2 \eps^{Br}_{hcm}} +
f_b \frac{\eps_b - \eps^{Br}_{hcm}}{\eps_b + 2 \eps^{Br}_{hcm}} = 0,
\l{Br_eq}
\end{equation}
which is  nonlinear in $\eps^{Br}_{hcm}$. A straightforward
manipulation of \r{Br_eq} delivers the explicit formula
\begin{equation}
\eps_a = \frac{\le f_a - 2 f_b \ri \eps_b + 2 \eps^{Br}_{hcm}}{ f_b
\le \eps_b - \eps^{Br}_{hcm} \ri + f_a \le \eps_b +2 \eps^{Br}_{hcm}
\ri} \, \eps^{Br}_{hcm} \l{ea_eq}
\end{equation}
 for $\eps_a$ in
terms of $\eps_b$, $\eps^{Br}_{hcm}$, $f_a$ and $f_b$. Since the
component materials $a$ and $b$ are treated in an identical manner
within the Bruggeman formalism, the corresponding formula for
$\eps_b$ has the same form as \r{ea_eq}. Notice that as the inverse
Bruggeman equation \r{ea_eq} does not involve a square root, there
is no scope for $\mbox{Im} \lec \eps_a \ric$ being nonzero if
$\eps_b, \eps^{Br}_{hcm} \in \mathbb{R}$. This contrasts with the
forward Bruggeman formalism where a square root term enables
$\mbox{Im} \lec \eps^{Br}_{hcm} \ric $ to be nonzero even though
$\eps_a, \eps_b \in \mathbb{R}$. This physically--implausible
scenario can arise when $\eps_a / \eps_b < 0$ \c{ML_OC1}.

For comparison, we introduce the Maxwell Garnett estimate of the HCM
relative permittivity \c{Ward}
\begin{equation}
\eps^{MG}_{hcm} = \eps_b + \frac{3 f_a \eps_b \le \eps_a - \eps_b
\ri}{\eps_a + 2 \eps_b - f_a \le \eps_a - \eps_b \ri } \l{Eq_MG}
\end{equation}
 and its corresponding inverse
\begin{equation}
\eps_a = \frac{\le 2 + f_a \ri \eps^{MG}_{hcm} - 2 f_b \eps_b}{\le 1
+ 2 f_a \ri \eps_b - f_b \eps^{MG}_{hcm}}\, \eps_b. \l{ea_mg}
\end{equation}
The limiting behaviour of the inverse Bruggeman estimate \r{ea_eq}
as compared with that of the inverse Maxwell Garnett estimate
\r{ea_mg} is especially revealing. In the limit $f_a \to 1$, both
estimates yield the relative permittivity of the HCM, as they
must.\footnote{The Maxwell Garnett estimate of the HCM relative
permittivity  is only strictly applicable in the dilute composite
regime $f_a \lesssim 0.3$. Accordingly, estimates of $\eps_a$
delivered by the inverse Maxwell Garnett formalism are strictly
valid  only for $f_a \lesssim 0.3$. However, $\eps^{MG}_{hcm}$
coincides with one of the Hashin--Shtrikman bounds on the HCM
relative permittivity which applies at all values of $f_a$ \c{H-S}.}
In the limit $f_a \to 0$, the inverse Bruggeman formalism yields
$\eps_a \to - 2 \eps^{Br}_{hcm}$ whereas the inverse Maxwell Garnett
formalism yields $\eps_a \to - 2 \eps_b$. Therefore, the two inverse
estimates differ markedly as $f_a$ approaches zero, provided that
$\eps_b$ and the relative permittivity of the HCM are sufficiently
different.

We now explore the inverse Bruggeman estimate \r{ea_eq}, in
comparison with the inverse Maxwell Garnett estimate \r{ea_mg}, by
means of some illustrative numerical examples. For  nondissipative
scenarios, the forward Bruggeman formalism runs into difficulties
when $\eps_a / \eps_b < 0$  but not when $\eps_a / \eps_b > 0$
\c{ML_OC1}. Accordingly, let us begin by focussing on the regimes
$\eps^{Br,MG}_{hcm} / \eps_b < 0$  and $\eps^{Br,MG}_{hcm} / \eps_b
> 0$. In Fig.~\ref{fig1}, plots of $\eps_a$, as determined by the
inverse Bruggeman formalism and the inverse Maxwell Garnett
formalism, versus $f_a $ are provided for the cases where $\eps_b =
\pm 2$ and $\eps^{Br,MG}_{hcm} = 3$. When $\eps^{Br,MG}_{hcm} /
\eps_b
> 0$ the inverse Bruggeman and inverse Maxwell Garnett estimates are
in fairly close agreement. However, the values of $\eps_a$ yielded
by  the two inverse formalisms differ markedly when
$\eps^{Br,MG}_{hcm} / \eps_b < 0$, except in the limit as $f_a$
approaches unity. Most notably, the inverse Bruggeman estimate
becomes singular and undergoes a change in sign as the volume
fraction increases  through $f_a = 0.56$, whereas the inverse
Maxwell Garnett value remains finite and does not change sign.

Next we turn to dissipative homogenization scenarios. In the case of
the forward Bruggeman formalism,  problems arise when $\mbox{Re}
\lec \eps_a \ric / \mbox{Re} \lec \eps_b \ric < 0 $ and the degree
of dissipation is relatively small; if $\mbox{Re} \lec \eps_a \ric /
\mbox{Re} \lec \eps_b \ric < 0 $ and the degree of dissipation is
relatively large or  if $\mbox{Re} \lec \eps_a \ric / \mbox{Re} \lec
\eps_b \ric > 0 $ then the forward Bruggeman formalism was found to
deliver physically plausible estimates of the HCM relative
permittivity  \c{ML_OC1}. Accordingly, we consider the regimes where
$\mbox{Re} \lec \eps^{Br,MG}_{hcm} \ric / \mbox{Re} \lec \eps_b \ric
> 0 $ with the degree of dissipation in the HCM being relatively
small, moderate and large. Graphs of the real and imaginary parts of
$\eps_a$, as estimated by the inverse Bruggeman and inverse Maxwell
Garnett formalisms, are plotted versus $f_a $ in Fig.~\ref{fig2} for
the cases  $\eps_b =  2$ and $\eps^{Br,MG}_{hcm} = 3 + \delta i$
where $\delta \in \lec 0.1, 1, 10 \ric$. When the degree of HCM
dissipation is relatively small ($\delta = 0.1$), the estimates of
the real and imaginary parts of $\eps_a$ provided by the inverse
Bruggeman and inverse Maxwell Garnett formalisms agree fairly
closely. When the degree of HCM dissipation is moderate ($\delta =
1$),  there is still fairly close agreement between the inverse
Bruggeman and inverse Maxwell Garnett values of $\eps_a$ for most
values of $f_a$. Crucially, however, for $f_a < 0.05$ the imaginary
part of $\eps_a$ estimated by the inverse Bruggeman formalism is
negative--valued (unlike $\mbox{Im} \lec \eps_a \ric$ estimated by
the inverse Maxwell Garnett formalism which is positive--valued).
Here $\mbox{Im} \lec \eps_a \ric < 0$ is not a physically plausible
outcome as it implies that the homogenization of an active material
$a$ and a nondissipative material $b$ results in a dissipative HCM.
For both the  real and imaginary parts of $\eps_a$, the
discrepancies between
 the values estimated by the two
inverse formalisms become enormous when the degree of HCM
dissipation is relatively large ($\delta = 10$). Furthermore, the
inverse Bruggeman estimate is  physically implausible for a much
larger range of $f_a$ values; i.e., $\mbox{Im} \lec \eps_a \ric$
estimated by inverse Bruggeman formalism is negative--valued for
$f_a < 0.3$ when $\delta = 10$.

Lastly, we explore the  $\mbox{Re} \lec \eps^{Br,MG}_{hcm} \ric /
\mbox{Re} \lec \eps_b \ric < 0 $ regime. Plots of the real and
imaginary values of $\eps_a$ in Fig.~\ref{fig3} correspond to  the
same  parameter values as those used for Fig.~\ref{fig2} except that
here $\eps_b = -2$. The estimates of the inverse Bruggeman formalism
are now physically implausible~---~due to $\mbox{Im} \lec \eps_a
\ric < 0$~---~for a wide range of $f_a$ values, regardless of
whether  the degree of HCM dissipation is relatively small, moderate
or large. In contrast, the estimate of
 $\mbox{Im} \lec \eps_a \ric $ provided by  the inverse
Maxwell Garnett formalism is positive--valued for all scenarios
considered. Additionally, the real parts of $\eps_a$ delivered by
the two inverse formalisms  differ enormously except when $f_a $
approaches unity, for all degrees of HCM dissipation considered.

\section{Closing remarks}

In the case of dissipative HCMs, the inverse Bruggeman estimates of
$\eps_a$ can be physically implausible when:
\begin{itemize}
\item[(i)] $\mbox{Re} \lec \eps^{Br}_{hcm} \ric / \mbox{Re} \lec
\eps_b \ric > 0 $ and the degree of HCM dissipation is moderate or
greater; or \item[(ii)]  $\mbox{Re} \lec \eps^{Br}_{hcm} \ric /$ $
\mbox{Re} \lec \eps_b \ric$ $ < 0 $ regardless of the degree of HCM
dissipation.
\end{itemize}
In the case of nondissipative HCMs, enormous discrepancies can exist
between the estimates of $\eps_a$  provided by the inverse Bruggeman
formalism and the inverse Maxwell Garnett formalism when
$\eps^{Br,MG}_{hcm} / \eps_b < 0$. Therefore, the inverse Bruggeman
formalism should be applied with great caution. Finally, we note
that in the very recent implementations of the inverse Bruggeman
formalism which motivated this study \c{ML_JNP,ML_PJ,ML_PNFA}, the
relative permittivity parameters were positive--valued and the
materials were nondissipative. The estimates yielded by the inverse
Bruggeman formalism in these cases seem physically plausible, but
the acid test can only be provided by suitable experimental
measurements.

\vspace{10mm}

\noindent {\bf Acknowledgment:} TGM is supported by a  Royal Academy
of Engineering/Leverhulme Trust Senior Research Fellowship.

\newpage

\begin{figure}[!ht]
\centering
\includegraphics[width=2.9in]{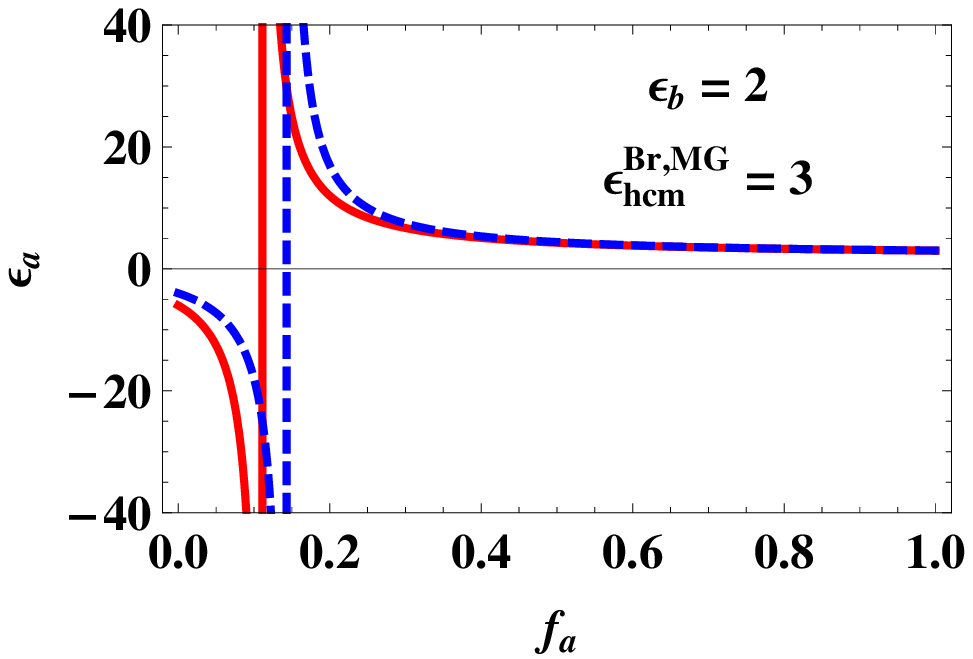}
\includegraphics[width=2.9in]{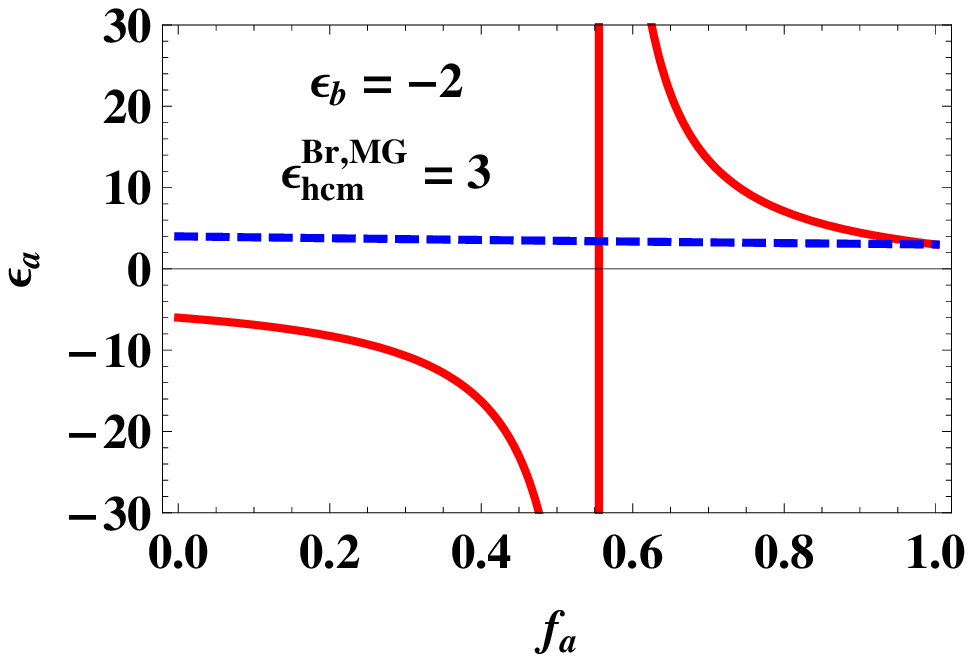}
 \caption{\l{fig1}
Plots of $\eps_a$ as determined by the inverse Bruggeman formalism
(red, solid curves) and the inverse Maxwell Garnett formalism (blue,
dashed curves) versus $f_a$ for
 $\eps_b = \pm 2$ and $\eps^{Br,MG}_{hcm} = 3$.
Estimates of $\eps_a$ delivered by the inverse Maxwell Garnett
formalism are strictly valid only for  $f_a \lesssim 0.3$.}
\end{figure}

\vspace{20mm}

\newpage

\begin{figure}[!ht]
\centering
\includegraphics[width=2.9in]{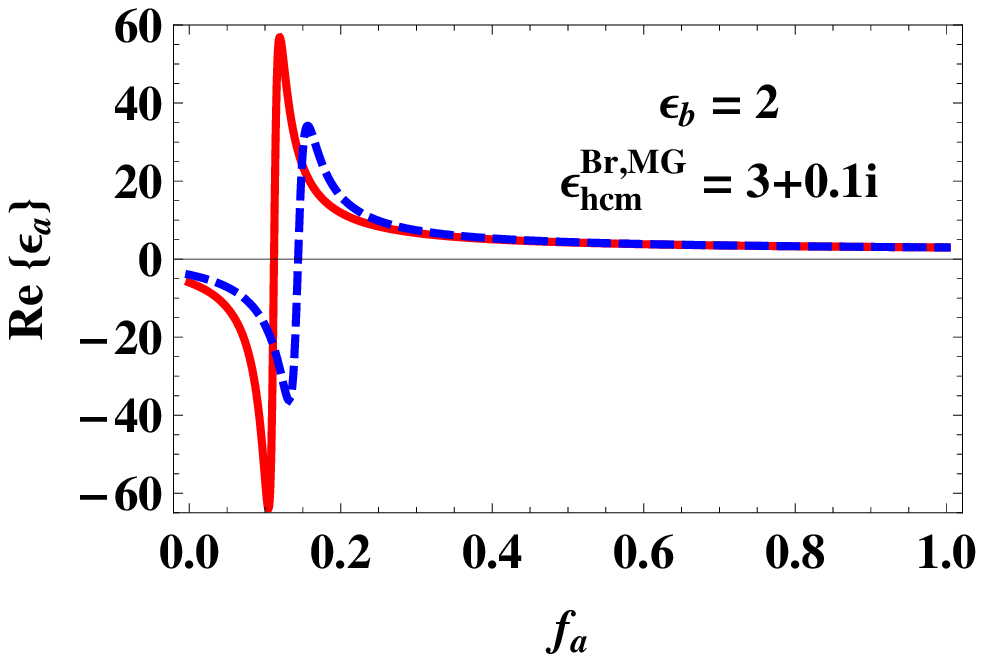}
\includegraphics[width=2.9in]{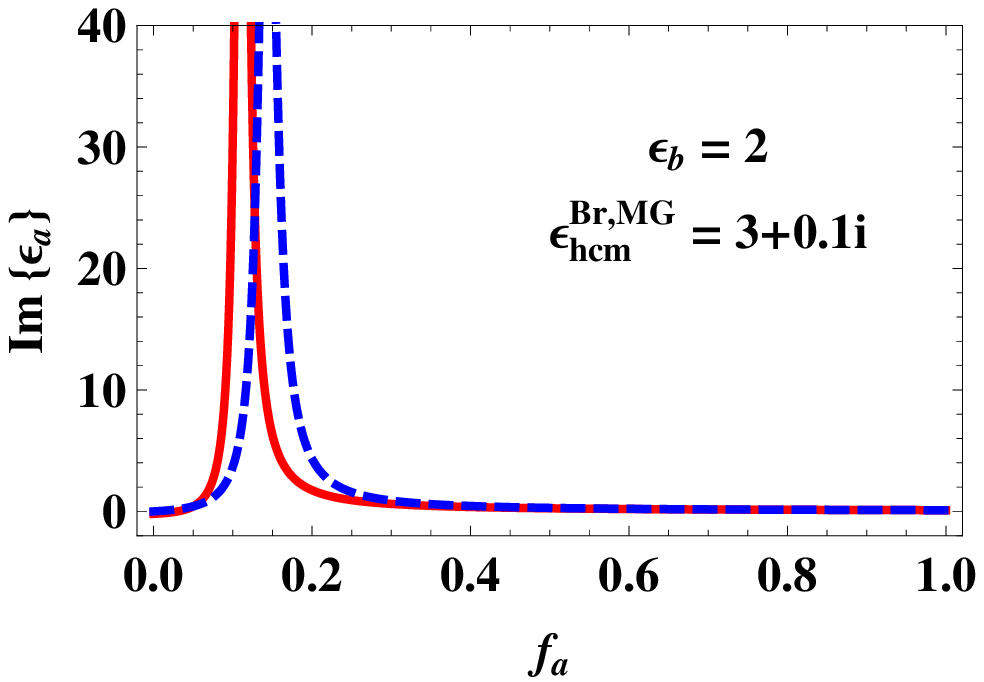}\\
\includegraphics[width=2.9in]{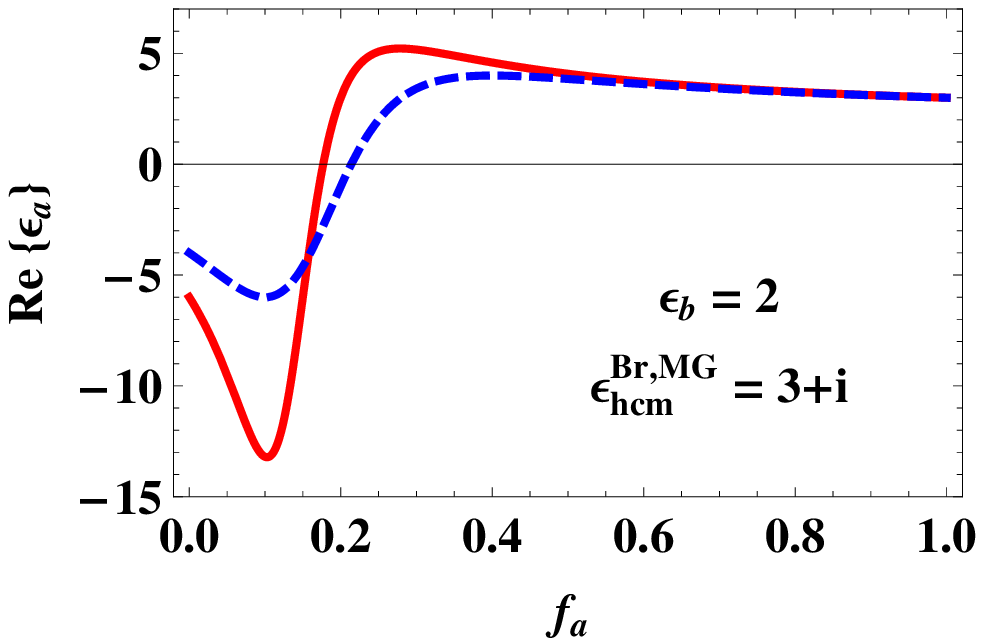}
\includegraphics[width=2.9in]{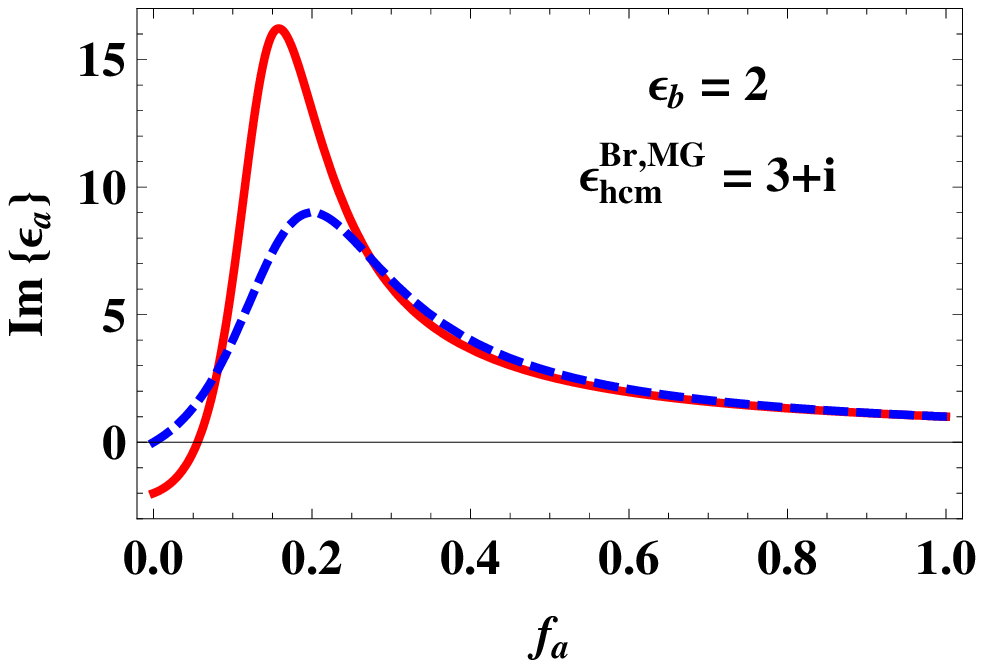}\\
\includegraphics[width=2.9in]{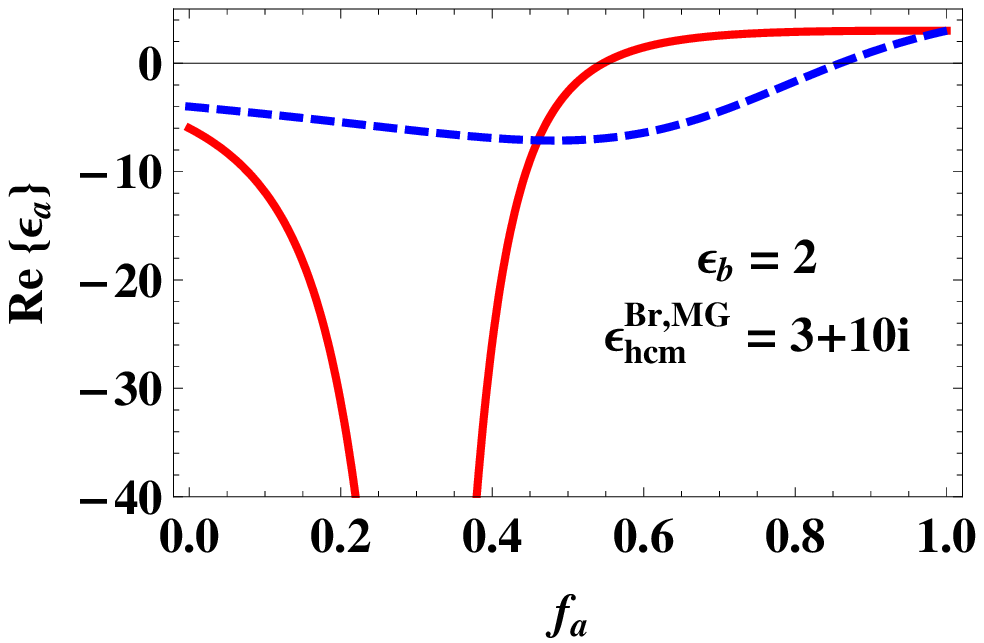}
\includegraphics[width=2.9in]{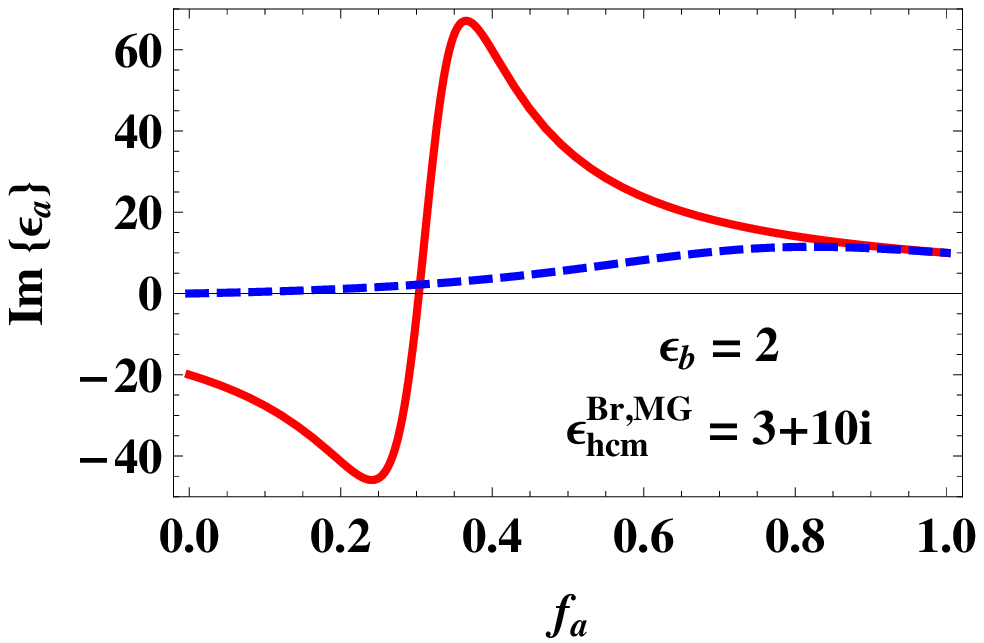}
 \caption{\l{fig2}
Plots of the real and imaginary parts of  $\eps_a$ as determined by
the inverse Bruggeman formalism (red, solid curves) and the inverse
Maxwell Garnett formalism (blue, dashed curves) versus $f_a$ for
 $\eps_b =  2$ and
$\eps^{Br,MG}_{hcm} = 3 + \delta i$ where $\delta \in \lec, 0.1, 1,
10 \ric$. Estimates of $\eps_a$ delivered by the inverse Maxwell
Garnett formalism are strictly valid only for  $f_a \lesssim 0.3$.
 }
\end{figure}

\newpage

\begin{figure}[!ht]
\centering
\includegraphics[width=2.9in]{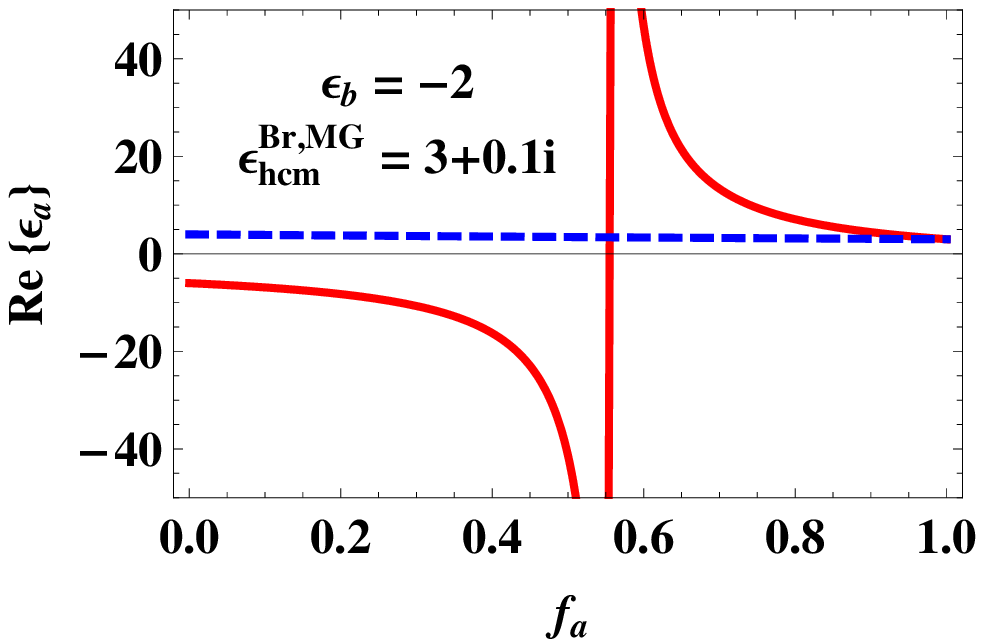}
\includegraphics[width=2.9in]{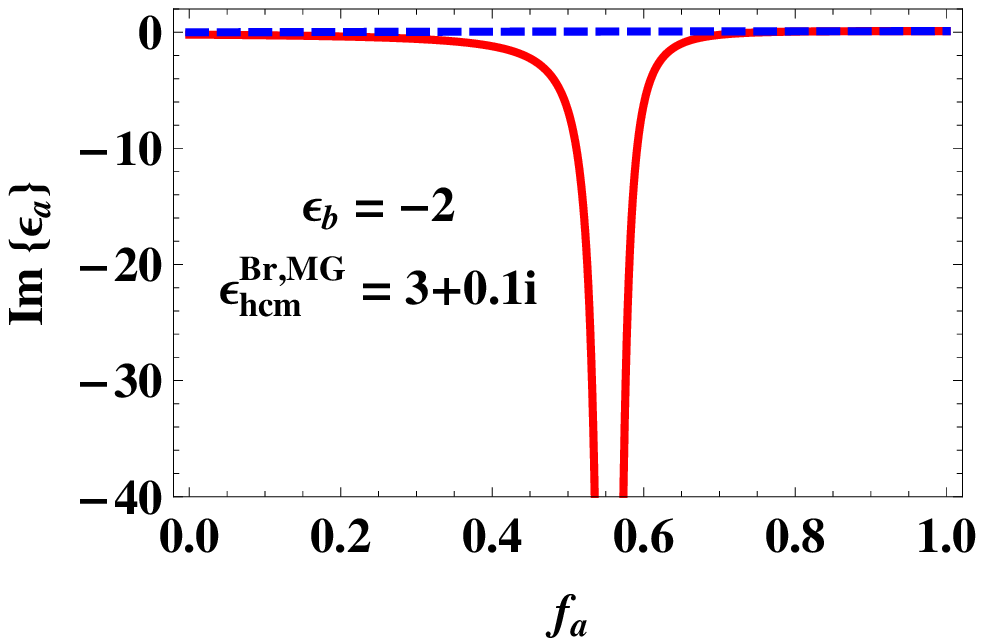}\\
\includegraphics[width=2.9in]{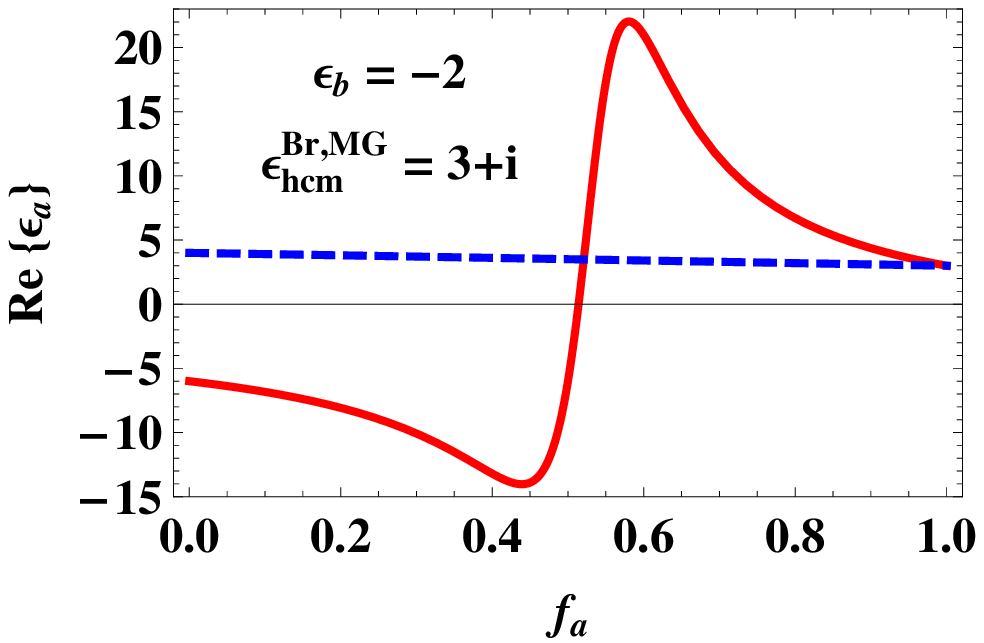}
\includegraphics[width=2.9in]{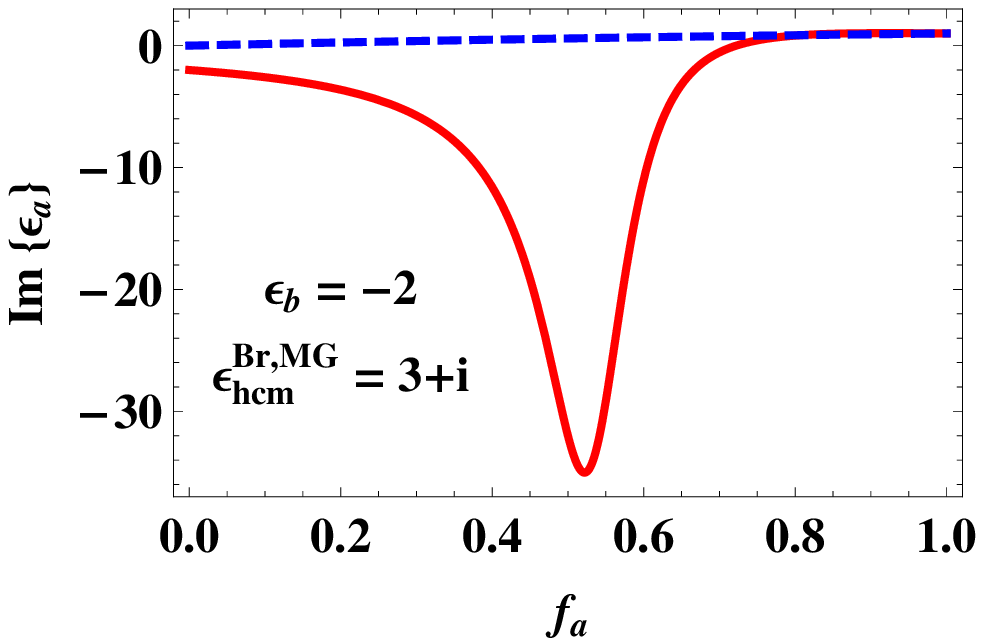}\\
\includegraphics[width=2.9in]{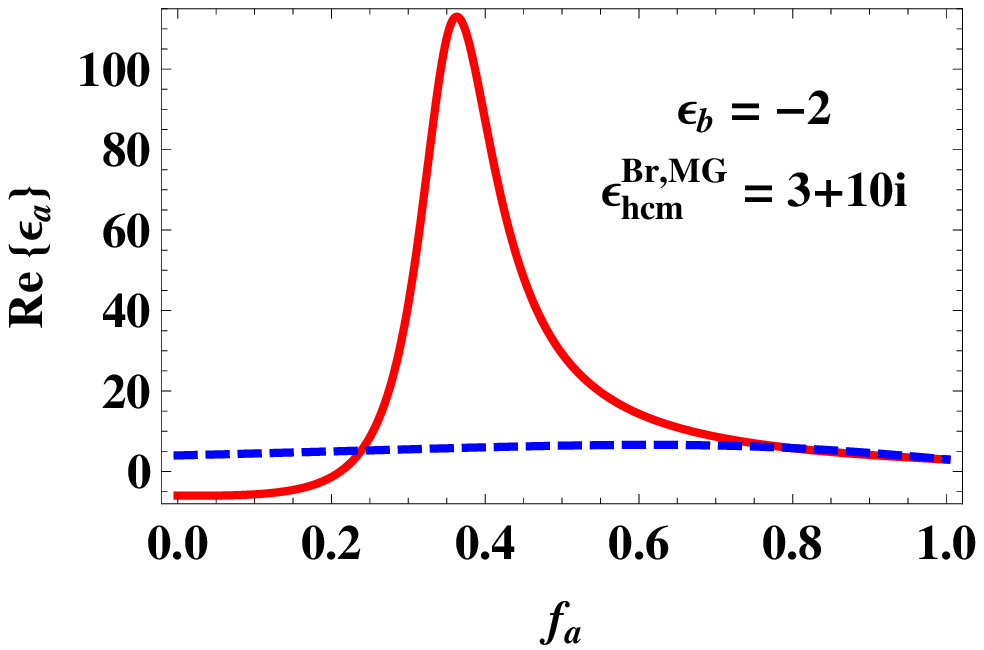}
\includegraphics[width=2.9in]{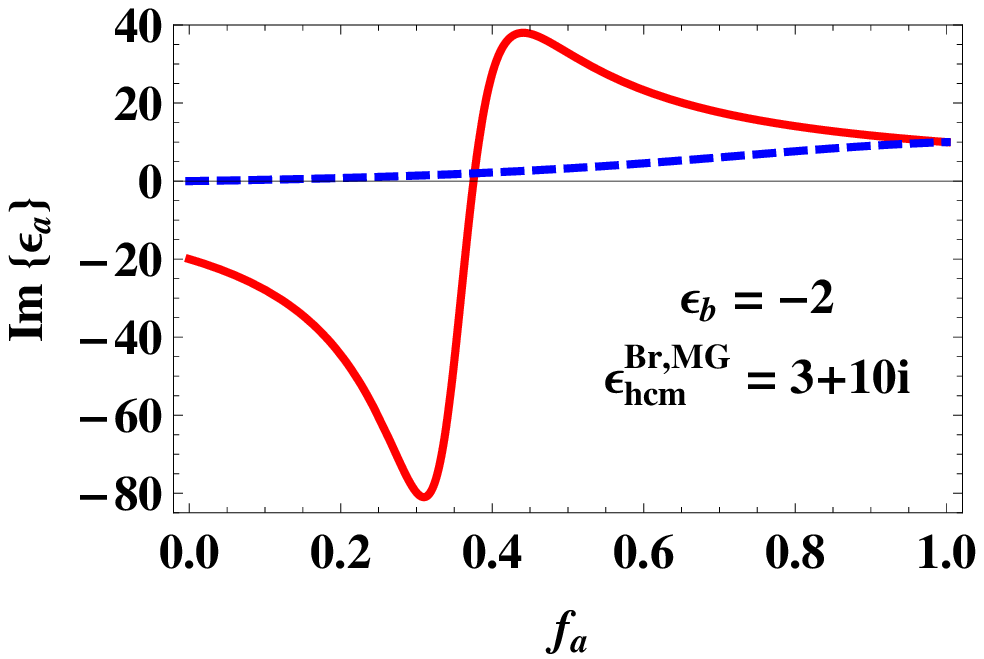}
 \caption{\l{fig3}
As Fig.~\ref{fig2} except that  $\eps_b = -2$.
 }
\end{figure}


\vspace{10mm}


\begin{thebibliography}{99}


\bibitem{L96} A. Lakhtakia  (ed),
Selected papers on linear optical composite materials, SPIE Optical
Engineering Press, Bellingham, WA, USA, 1996.


\bibitem{M_electromagnetics}
 T.G. Mackay,
 Linear and nonlinear homogenized composite mediums as metamaterials,
Electromagnetics  25 (2005), 461--481.


% \bibitem{Br}
% D.A.G. Bruggeman,   Berechnung verschiedener physikalischer
%Konstanten von heterogenen Substanzen, I.
%Dielektrizit\"ats\-konstanten und Leit\-f\"ahig\-keiten der
%Misch\-k\"orper aus isotropen Substanzen,  Ann Phys Lpz  24 (1935),
%636--679. (Reproduced in \c{L96}).


\bibitem{Ward} L. Ward,  The optical constants
of bulk materials and films, 2nd ed, IOP, Bristol, UK, 2000.

\bibitem{WM_AEU}
W.S. Weiglhofer and T.G. Mackay,  Numerical studies of the
constitutive parameters of a chiroplasma composite medium, Arch
Elektron \"Ubertrag~---~Int J Electron  Commun  54 (2000), 259--265.

\bibitem{MW_JOPA}
T.G. Mackay and W.S. Weiglhofer, Homogenization of biaxial composite
materials: dissipative anisotropic properties, J Opt A:  Pure Appl
Opt 2 (2000), 426--432.

\bibitem{ML_NPV_MOTL}
T.G. Mackay and A. Lakhtakia,  Negative phase velocity in isotropic
dielectric-magnetic media via homogenization, Microwave Opt Technol
Lett 47 (2005), 313--315.


\bibitem{ML_OC1}
 T.G. Mackay and A. Lakhtakia, A limitation
of the Bruggeman formalism for homogenization, Opt Commun  234
(2004), 35--42. Erratum  282 (2009), 4028.


\bibitem{M_Ag}
 T.G. Mackay,
 On the effective permittivity of silver--insulator
nanocomposites, J Nanophoton 1 (2007),  019501.


\bibitem{ML_OC2}
 T.G. Mackay and  A. Lakhtakia,
 On the application of  homogenization formalisms to active dielectric
composite materials, Opt Commun  282 (2009), 2470--2475.

\bibitem{Aspnes} D.E. Aspnes,  Local--field
effects and effective--medium theory: a microscopic perspective, Am
J Phys  50 (1982), 704--709. (Reproduced in \c{L96}).


\bibitem{Duncan}
A.J. Duncan,   T.G. Mackay, and  A. Lakhtakia,
 On the Bergman--Milton bounds for the homogenization of
 dielectric composite materials,
 Opt Commun  271 (2007), 470--474.



\bibitem{Bross1}
A. Mejdoubi and C. Brosseau, Intrinsic electrostatic resonances of
heterostructures with negative permittivity from finite-element
calculations: Application to core-shell inclusions, J Appl Phys 102
(2007), 094104.

\bibitem{Bross2}
C. Fourn and C. Brosseau, Electrostatic resonances of
heterostructures with negative permittivity: Homogenization
formalisms versus finite-element modeling,  Phys Rev E 77  (2008),
016603.

\bibitem{Bross3}
A. Mejdoubi and C. Brosseau, Controlling intrinsic electrostatic
resonances of negative permittivity artificial multilayers, J Appl
Phys 103  (2008), 084115.



\bibitem{WSW_MOTL}
W.S. Weiglhofer, On the inverse homogenization problem of linear
composite materials, Microwave Opt Technol Lett 28 (2001), 421--423.

\bibitem{Cherkaev}
E. Cherkaev, Inverse homogenization for evaluation of effective
properties of a mixture, Inverse Problems 17 (2001), 1203--1218.


\bibitem{ML_JNP}
 T.G. Mackay and  A. Lakhtakia,
 Determination of constitutive and
morphological parameters of columnar thin films by inverse
homogenization, J Nanophoton  4 (2010),  041535.


\bibitem{ML_PJ}
T.G. Mackay and  A. Lakhtakia,
 Empirical model of optical sensing via spectral shift
of circular Bragg phenomenon, IEEE Photonics J  2 (2010), 92--101.

\bibitem{ML_PNFA}
 T.G. Mackay and  A. Lakhtakia,
 Modeling  columnar thin films as platforms for
surface--plasmonic--polaritonic optical sensing, Photon Nanostruct
Fundam  Appl (at press). doi:10.1016/j.photonics.2010.02.003

\bibitem{H-S} Z. Hashin and S.  Shtrikman,
 A variational approach to the theory of the effective magnetic
permeability of multiphase materials, J Appl Phys  33 (1962),
3125--3131.

%\bibitem{MLW_SPFT}
%   T.G. Mackay, A. Lakhtakia and W.S.
%Weiglhofer, Strong-property-fluctuation theory for homogenization of
%bianisotropic composites: formulation, Phys Rev E  62 (2000),
%6052--6064. Erratum  63 (2001),  049901.


%\bibitem{L_nihility}

\end{thebibliography}
\end{document}